\documentclass[11pt,a4paper]{article}
\usepackage[left=2 cm, right=2 cm, top=2.5 cm, bottom=3 cm]{geometry} 
\usepackage[pdfborder={0 0 0},plainpages=false,pdfpagelabels]{hyperref}
\usepackage[latin1]{inputenc}
\usepackage{amsmath}
\usepackage{amsfonts}
\usepackage{amssymb}
\usepackage{makeidx}
\usepackage{graphicx}
\usepackage{cite}
\usepackage{multicol} 
\usepackage{multirow}
\usepackage{tabularx}
\usepackage{subfig}
\begin{document}
\begin{center}
\textsc{\Large \textbf{RESULTS OF A ONE-WAY EXPERIMENT TO TEST THE ISOTROPY OF THE SPEED OF LIGHT}}\\[0.5cm] 
Md. Farid Ahmed$^{1*}$, Brendan M. Quine$^{1,2}$, Spiros Pagiatakis$^{1,2}$ and A. D. Stauffer$^{2}$ \linebreak\linebreak
$^{1}$Department of Earth and Space Science and Space Engineering, York University, 4700 Keele Street, Toronto, Ontario, Canada-M3J 1P3. \linebreak\linebreak  
$^{2}$Department of Physics and Astronomy, York University, 4700 Keele Street, Toronto, Ontario, Canada-M3J 1P3.
\\[0.5cm] 
$^{*}$Corresponding author: E-mail: mdfarid@yorku.ca \\[0.4cm]
\end{center}
\bigskip
\textbf{ABSTRACT:}
This paper presents the outcome of an experiment based on an improved version of Fizeau's coupled-slotted-
discs that tests the fundamental postulates of Special Relativity for the one-way speed of light propagation. According to our methodology, important phenomena - a limit on and the diurnal regularity of the variation of the speed of light due to the movements of the Earth (assuming that the speed of light follows a Galilean transformation) - can be tested by the present experiment. However, these measurements do not indicate any significant diurnal variation. Consequently, the limit of the present outcome on the variation of the speed of light is insignificant. Assuming that the speed of light is not invariant and performing a rigorous statistical analysis, the limit established is approximately 1/50 of the previous Fizeau-type experiment with $95\%$ confidence level. These outcomes are consistent with the assumptions of Einstein's Special Relativity. 
\bigskip\\
\textbf{Shortened version of the title:}\\
RESULTS OF A ONE-WAY ISOTROPY TEST 
\bigskip\\
\section{Introduction} Special Relativity (SR) helps to unify electrodynamics with mechanics and it became the cornerstone of Modern Physics. SR constrains our conceptions of time, space and the existence of a preferred cosmological reference frame. Consequently, this phenomenon has been subject to considerable experimental scrutiny since its birth which make SR the most acceptable and reliable theory of motion to date. Despite all these successes, the fundamental postulate of SR is facing a new challenge by quantum theories of gravity \cite{KS1, AME2, GP3}. Any deviation from this postulate would help physicists unify all fundamental forces in nature. Therefore this fundamental postulate has been the subject of a variety of experimental tests \cite{CH4}. \\ \\ 
According to the Robertson-Mansoury-Sexl (RMS) \cite{RB5, MS6, MS7, MS8} test theory, experiments that test the isotropy of the speed of light may be divided into two types based on the propagation of lights: (a) one-way (single-trip or first order), and (b) two-way (round-trip averaged or second order). Most of these tests utilize a two-way methodology which can establish only the second order or round-trip averaged speed of light over closed paths \cite{KM9, AB10}. However, these approaches provide no experimental results for the one-way isotropy of the speed of light which is still unresolved \cite{AB10, SH11, ED12, AB13}.  \\ \\
According to Will \cite{WI14}, the consideration of clock synchronization is irrelevant for a one-way speed of light test if one expresses the results in terms of physically measurable quantities. Spavieri  \cite{SP15, SP16} showed that it is possible to measure a one-way speed of light. Reports \cite{MT17, RA18, NH20} stress the need to repeat the measurements performed by Marinov \cite{MA21, MA22, MA23} who used an improved version of the Fizea-type-coupled-slotted-discs apparatus and reported a controversial result \cite{AB10, MA24}. Despite all these proposals, discussions and criticisms, there has been no repetition so far. We performed the experiment. Detailed descriptions of the experimental apparatus, different challenges and an interpretation of the present experiment were presented elsewhere \cite{AB13}. We present the results in this report.
\section{An interpretation}
The experiment consists of a shaft with two holed discs mounted at its ends and rotated by an electromotor. Light from a source is divided into two beams by a beam-splitter and several adjustable mirrors that are directed to the opposite ends of the rotating shaft, so that the light rays can pass through the holes of the discs in mutually opposite directions and illuminate two photo detectors. The two beams are chopped by the near disc and have to transit the undergoing length to reach the far disc.\\ \\
The responses of the photo detectors are collected at a motor speed $N$ RPM. The average of these collected responses from a detector D$_{1} $ is called the response $R_{zi} $ for the light propagation along $z$-axis for any Run (a sample) $i$ in an event. Similarly, the average of the collected responses from a detector D$_{2} $ is called the response $ R_{-zi} $ for the light propagation along $-z$-axis. There is a series of Runs $(1 .. i, i+1, ..)$ during the course of an event for a 24-hour period during which the Earth makes a complete rotation. We derived the time dependent component of the velocity $\mathbf{v}(t)$ of the laboratory relative to a reference frame along the direction of the light propagation in the paper \cite{AB10}. Therefore, according to a Galilean transformation, if the speed of light depends on the speed of the source or the speed of the observer then the responses should be proportional to the variation of $\mathbf{v}(t)$ during the course of an event for a 24-hour period. \\ \\
It is challenging to construct two duplicate detectors having exactly the same apertures as well as two duplicate discs having exactly the same size of slits to the desired accuracy for an apparatus. Consequently, it is challenging to generate the responses of the detectors exactly identical. Note that we need to compare two detectors' responses which are identical due to construction to the desired accuracy. In order to avoid these challenges, we normalize the responses. Following the reports \cite{AB10, AB13}, we can derive the normalized responses $ R_{zi\_N} $, $R_{-zi\_N}  $ and $ R_{diff\_i\_N} $ due to light propagations along $ z $- axis, $ -z $-axis and their differential, respectively, as follows:
\begin{eqnarray}
\label{eq:northnormalized1}
R_{zi\_N}=&\frac{R_{zi}}{\overline{R}_{z}} \approx 1+ \frac{v_{1}(t)}{c_{0}} +O \left( \frac{1}{c_{0}^{2}}\right)+\cdots \\  
\label{eq:northnormalized2}     
R_{-zi\_N}=&\frac{R_{-zi}}{\overline{R}_{-z}} \approx 1- \frac{v_{1}(t)}{c_{0}} +O \left( \frac{1}{c_{0}^{2}}\right) +\cdots \\    
\label{eq:northnormalized3}   
R_{diff\_i\_N} =&R_{zi\_N}-R_{-zi\_N}\approx 2 \frac{v_{1}(t)}{c_{0}}      
\end{eqnarray}
where, $\overline{R}_{z}$ and $ \overline{R}_{-z} $ are grand means of $R_{zi}$ and $ R_{-zi} $ respectively for the course of an event for a 24-hour period. Also, $ v(t)=v_{0}+v_{1}(t) $ where $v_{0}$ is a constant term which is the component along Earth's rotation axis and $v_{1}(t)$ is the sinusoidal term which is perpendicular to the axis. These first order effects derived in equations (\ref{eq:northnormalized1}), (\ref{eq:northnormalized2}) and (\ref{eq:northnormalized3}) are being tested with an experiment described in the present work.  \\ \\
Ideally, according to our experimental procedure, we assume two identical responses for the opposite detectors. Then we look for any change of 24-hour period keeping that ideal condition throughout the measurement. Therefore, according to our predictions above, we can assume that all maxima and minima of the responses (north, south, and difference) appear at the same time of the day of measurement as illustrated in Fig. \ref{fig:sriv310}.
\\ \\ \\
\noindent%
\begin{minipage}{\linewidth}
\makebox[\linewidth]{
  \includegraphics[keepaspectratio=true,scale=0.5]{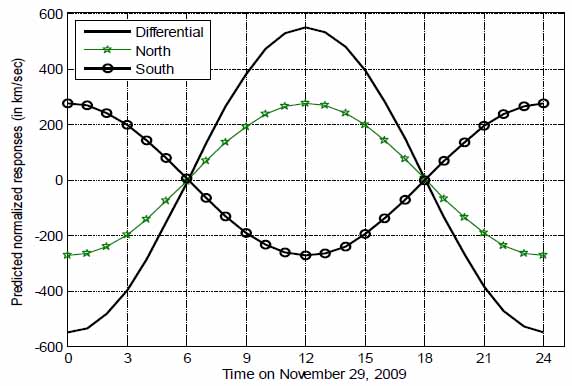}}
\captionof{figure}{An illustration of predicted relative first order effects if SR is invalid.}
\label{fig:sriv310}
\end{minipage}
\\ \\
\section{Description of the present test}
The experimental apparatus is illustrated in Fig. \ref{figure:setupmodern43a}. An NL-1 ultra-stable \cite{IN25} frequency stabilized He-Ne Laser, (Newport) (Wavelength $= 632.8$ nm; Frequency $= 474 \times 10^{6}$ MHz; Output Power $ >0.7 $ mW; Transverse Mode: TEM$_{00} $; Beam Diameter at exit at $ e^{-2} $ intensity$=0.63$ mm) is used in these measurements. Two orthogonally-polarized modes (H and V) oscillate simultaneously in the laser gain tube. The laser beam passes through a polarizing cube beam splitter PBS, which diverts one mode (H) to illuminate a photodiode D4. The polarized beam (V) passes through a non-polarizing dielectric beam splitter BS1, which diverts about $20\%$ of the beam to another photodiode D3. The outputs of the two photodiodes (D3 and D4) - representing the intensities of the two polarizations modes in the raw beam - are monitored by a differential comparator circuit which drives a thermal servo mechanism that keeps the laser cavity tuned to a particular resonance frequency \cite{IN25}. These He-Ne lasers were studied in the laboratory at the Joint Institute for Laboratory Astrophysics (JILA). They reported the frequency stability was better than 1 part in $ 10^{10} $ over the periods of about 1 hour and 1 part in $ 10^{8} $ over 1 year \cite{NI26}. Also the Allan variance \cite{AL27} plot for a typical red side lock for He-Ne laser was presented in the same report \cite{NI26}. However, environmental conditions must be stabilized in order to get a stable laser. The experiment operates for an extended period of at least 24 hours for one run data set. We monitor temperatures (T1, T2, T3 and T4), humidity and atmospheric pressure in the laboratory. 
\\ \\ 
\noindent%
\begin{minipage}{\linewidth}
\makebox[\linewidth]{%
  \includegraphics[keepaspectratio=true,scale=0.6]{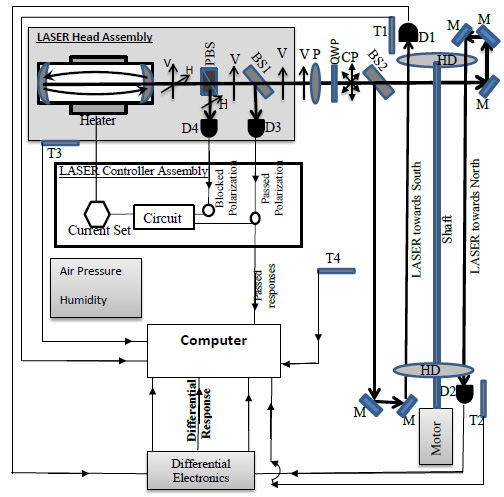}}
\captionof{figure}{Block Diagram of the improved experimental setup. D1 to D4 are Photodiodes; T1 to T4 are Temperature sensors; M is Mirror;  HD is Holed Disc; BS1 and BS2 are Beam Splitters; QWP is Quarter Wave Plate; PBS is Polarizing cube Beam Splitter; V (Vertical), H (Horizontal) are Polarization mode; P is Polarizer; CP is Circular Polarization.}
\label{figure:setupmodern43a}
\end{minipage}
\\ \\
For our improved apparatus, we collect the passed response $ P_{i} $ which is the average of the collected responses from a detector D$_{3} $ at the same time as $ R_{zi} $ and $ R_{-zi} $. Changes in internal laser intensity due to external variations of environmental conditions can be identified using these passed responses.\\ \\
Both long- and short-term stability can be maintained more easily by reducing vibration effects on critical components. Therefore, the Laser is mounted on a vibration-isolated table to reduce mechanical drift. Also, short-term stability is highly dependent on the amount of optical feedback \cite{CI28, DI29}, which is reduced by adjusting the polarizer P. Also, in order to convert linear polarization (V) into Circular Polarization (CP) a Quarter Wave Plate QWP is used. This QWP requires proper orientation to get maximum circularly polarized light. We rotate the polarizer using a motor for any orientation of the QWP and observe the signal using a Power meter. In the optimized orientation of the QWP, the signal response in the power meter is almost constant. The orientation of the QWP is fixed for the rest of the experiment. Circular polarization of the laser reduces intensity variations due to mirror reflections in the optical path. \\ \\
The laser beam is divided into two beams by a beam-splitter BS2 and adjustable mirrors M direct the beams to opposite ends of the rotating shaft, so that the beams pass through the holes of the discs in mutually opposite directions and illuminate photodiodes D1, D2. For our present experiment we used low noise silicon PIN photodiodes (Type OPF470, OPTEK, USA) and a precision high-speed transimpedance amplifier (Type OPA380, TEXAS INSTRUMENTS, USA). Also, the parameters were used: the aperture of the detector thorough which laser beam is entering, $d=(2.0\pm 0.05)$ mm; width of the slit, $d_{1}=(5\pm 0.05)$ mm; distance of the chopping point from centre of the disc, $R=(80\pm0.05)$ mm and the distance between two discs is 1.755 m. After observing and aligning both signals using an oscilloscope, we turn on our differential electronics that can converts photocurrent into potential difference and calibrate the zero differential response level. The conversion factor from photocurrent into voltage for this electronic device is 0.15 $\mu$A into 1 V. Our setup records differentials of 0.01 mV. Consequently, current changes as small as $\approx$0.0015 nA can be measured. \\ \\
The edges of the slits of the rotating discs can cause scattering and diffraction effects. Therefore, we set our detectors in enclosures with shielding as described in the report \cite{AB13}. The paths of the laser beams received by the detectors are connected by the black plastic cylindrical tubes. These tubes have internal diameter which are equal to the aperture of the detectors and the lengths are $ \approx $10 mm which help us to reduce the scattering and diffraction disturbances. However, we tested the scattering and diffraction effects by changing the width of a laser beam for different discs (also, changing sizes, colors of the discs) and did not see any significant change. Laser beam narrowing or broadening by changing the aperture of a diaphragm for different positions of the diaphragm and observing in the oscilloscope which indicates no significant change in the outcome. Therefore, we conclude that the present setup - using the black discs with slits of dimension ($6 mm\times5 mm\times 0.5 mm$), a gap between two slits of 8 mm and the detectors inside the shielding boxes as described in the report \cite{AB13} - minimizes scattering and diffraction effects. \\ \\
For any precision measurement, especially to test anisotropy of the speed of light, the important uncertainty is due to temperature variation \cite{JO31}, consequently, errors of measurements due to temperature changes were considered carefully in this experiment. The observer's own body-heat or infra-red radiation can produce an effect on the test results \cite{JO31}. Also the changes of room temperature for a 24-hours period are significant for any isotropy test \cite{JO31}. Following our approach, the typical resulting differential signal is reliable for $ \leq2 $ minutes. Using the alignment method of Marinov \cite{MA23}, we did not detect any reliable data. Therefore, we proposed a different alignment as well as the interpretation of the present test as already described in \cite{AB13}. According to our approach, both holed discs open and close at the same time.     \\ \\
In response to a photon flux (or optical power) a photodiode generates a proportional electric current. This current generated in the device is a random quantity whose value fluctuates above and below its average value. These fluctuations, generally regarded as noise, are characterized by the standard deviation of the response \cite{MM32}. Different sources of noise that are inherent in the process of photodiodes can be identified as photon noise, photoelectron noise, gain noise, receiver circuit noise, dark current noise, etc. \cite{MM32}. In order to understand the origin of the noise (due to photodiode and receiver circuit), evaluate its extent, and take them into account for our present experiment, we made measurements when the laser beams did not illuminate the photodiode detectors but the complete electronic system was turned on. We assume these responses of photodiodes as dark current responses for our present setup. The subtraction processes cancel out most of the noise produced by the dark current in the differential responses. 
\section{Error analysis}
There are two events of measurements of 24-hours duration each, in Nov. 14 - 15 (Event 1) and 28 - 29, 2009 (Event 2). Each event consists of 12 samples, each taken every two hours. Each sample comprises about 74 measurements with a sampling interval of 1 second. All measurements are performed in air with motor speed $ (3200\pm 16 )$ RPM. All samples pass the chi-square goodness-of-fit test for normality at the $95\%$ confidence level. The distribution shapes, as shown in Fig. \ref{figure:ch2test}, give us an idea about the samples which represent unbiased samples. In order to calculate means and variances of the means of all the samples, we follow the methods by Herman \cite{HE33}. \\ \\
\noindent%
\begin{minipage}{\linewidth}
\makebox[\linewidth]{%
  \includegraphics[keepaspectratio=true,scale=0.45]{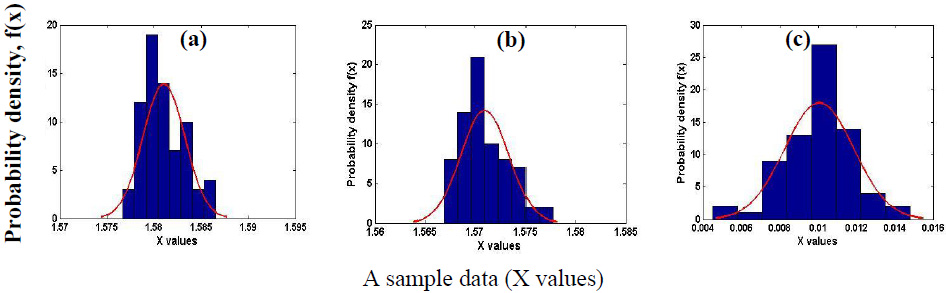}}
\captionof{figure}{Presents the histograms and normal probability distributions (f(X)) based on a sample data (X), of (a) the North, (b) the South and (c) the differential responses, for an event with sample size 74. Speed of the motor was (3200$ \pm $16) RPM and measurements were performed in air.}
\label{figure:ch2test}
\end{minipage}
\\ \\ \\
\noindent%
\begin{minipage}{\linewidth}
\makebox[\linewidth]{
  \includegraphics[keepaspectratio=true,scale=0.4]{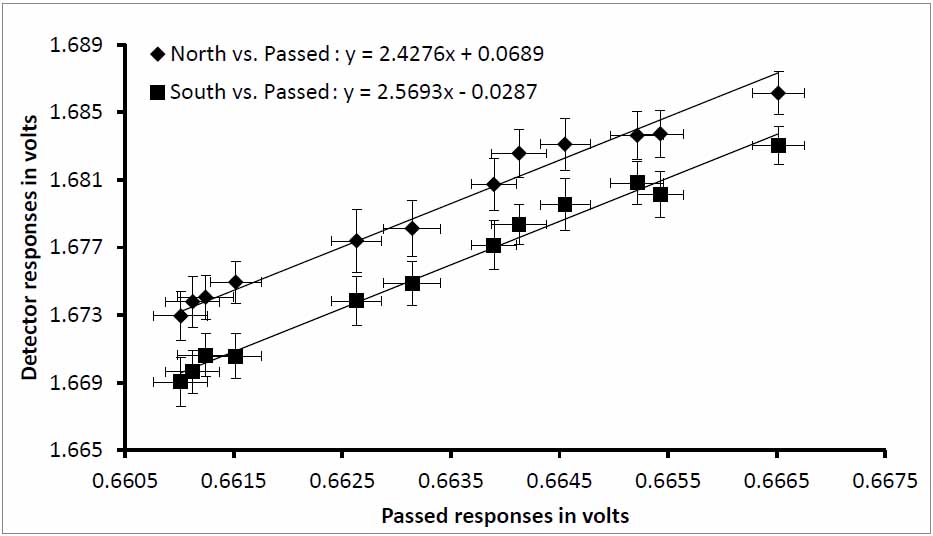}}
\captionof{figure}{An illustration of the relationship among north-responses, south-responses, and passed-responses in volts for a measurement with motor speed 3200 RPM. Note that we used the stabilized laser to make our final measurements. Error bars are 5$ \times $ amplified. }
\label{figure:nsvpr512}
\end{minipage}
\\ \\ \\
In order to minimize the long-term effect as shown in Fig. \ref{figure:nsvpr512}, we set out the change in a mathematical form. Note that we have used three detectors for our measurements: (i) north (N); (ii) south (S) and (iii) control (the passed responses). Let us consider $\Re_{N}$, $\Re_{S}$ and $\Re_{P}$ the responsivity of north, south and passed photodiodes, respectively. Following the report \cite{SP181}, we introduce the relation between the response of a detector $ \Re $ and the intensity of the laser $ I $ as follows:
\begin{equation}
\label{eq:relationpassin} 
\Re=AI+B
\end{equation}
where $ A $ is the slope and $ B $ is the constant term for this linear relation. \\ \\
In order to understand the variation due to laser intensity instability in the detectors' responses and to develop the relations between the passed-responses $ \Re_{P}=C_{1}I+C_{0} $ and detectors' responses ($\Re_{N}$, $\Re_{S}$) - following Fig. \ref{figure:nsvpr512} - let us consider, equations for the north and the south responses as follows:
\begin{eqnarray}
\label{eq:passnorth} 
\Re_{N} &=A_{N}I+B_{N}=A_{N}\left[ \dfrac{\Re_{P}-C_{0}}{C_{1}}\right] +B_{N}=\frac{A_{N}}{C_{1}}\Re_{P}+\left[ B_{N}-\frac{A_{N}C_{0}}{C_{1}}\right]  \\
\label{eq:passsouth}
\Re_{S} &=A_{S}I+B_{S}=A_{S}\left[ \dfrac{\Re_{P}-C_{0}}{C_{1}}\right] +B_{S}=\frac{A_{S}}{C_{1}}\Re_{P}+\left[ B_{S}-\frac{A_{S}C_{0}}{C_{1}}\right] 
\end{eqnarray}
where $ A_{N} $ is the slope for north response versus the passed-response, $ A_{S} $ is the slope for south response versus the passed-response, $ B_{N} $ and $ B_{S} $ are constants for the linear relations between the laser intensity and individual (north and south) responses. Eqs. (\ref{eq:passnorth}) and (\ref{eq:passsouth}) indicate that the intercepts of the nearly parallel lines as illustrated in Fig \ref{figure:nsvpr512} can be positive or negative depending on the arrangements of the devices (e.g. detectors' positions) in the apparatus. 
\\ \\
We use the stabilized laser to make final measurements. However, the relation, presented in Fig. \ref{figure:nsvpr512}, indicates that changes of the detectors' responses depend on that of the passed responses. Consequently, the north and south detectors' responses need to be corrected. According to our assumption, if special relativity is invalid (objective of the present test), the individual (N, S) responses depend on the Earth's velocity. Therefore, in order to model the two opposite responses (including variation due to the laser instability) when the discs are rotating, following French \cite{FR147}, let us consider a general equation assuming that the responses depend on the Earth's velocity $ v=v(t) $ and the passed response $ P_{i} $ as follows:
\begin{equation}
\label{eq:lnmodel}
\Re_{i} =A_{i}v(t)P_{i}+B_{i}
\end{equation}
where $ A_{i} $ and $ B_{i} $ are time independent constants. We introduce Eq. (\ref{eq:lnmodel}) to present the movements of the Earth (if SR is invalid) as well as the variation due to the laser instability that helps us to explain our steps for error analysis of the long-term effects.    
\\ \\
Following Brockwell and Davis \cite{BD180}, and the indication of an increasing trend with the passed responses which is apparent in both responses (N,S) as illustrated in Fig. \ref{figure:nsvpr512}, we can validate Eq. (\ref{eq:lnmodel}). Therefore, in order to gain a further understanding by analyzing our data, we follow the following steps:
\begin{enumerate}
  \item We plot the detectors' responses versus the passed-signal similar to Fig. \ref{figure:nsvpr512} and estimate the fitted equations accordingly to find the values of  $ B_{i} $. We use weighted least square fitting using a linear model.
  \item We calculate $ (\Re_{i}-B_{i}) $ following Eq. (\ref{eq:lnmodel}). Then, we perform normalization of $ (\Re_{i}-B_{i}) $ following the procedure as already explained in Section 2. These normalized responses can be denoted by $ [\Re_{i}-B_{i}]_{Nor} $.
  \item We calculate corrected responses of the individual detectors (the north and the south) using the passed-responses as follows:
  \begin{equation}
  \label{eq:srvalidinvalid2}
  \text{Corrected~response}=\frac{[\Re_{i}-B_{i}]_{Nor}}{P_{i}}
  \end{equation}
\item We calculate the differential responses using the individual responses as derived in Eq. (\ref{eq:srvalidinvalid2}). 
\end{enumerate}
\section{Results and discussions}
Following the procedures as described above, we perform error analysis of the individual (N, S) responses by the passed-responses of Events 1 and 2. Fig. \ref{figure:conindr1611}(a) presents the results of the individual responses (N, S) for Event 1. Similarly, Fig. \ref{figure:conindr2612}(a) presents the results for Event 2. The corresponding differential responses are presented in Figs. \ref{figure:conindr1611}(b) and \ref{figure:conindr2612}(b) respectively. Indeed, the variations in Figs. \ref{figure:conindr1611} and \ref{figure:conindr2612} do not appear to exhibit a diurnal variation following a Galilean transformation where the two opposite individual responses present opposite phases and the differential one doubles. This outcome is consistent with special theory of relativity.  \\ \\
\noindent%
\begin{minipage}{\linewidth}
\makebox[\linewidth]{
  \includegraphics[keepaspectratio=true,scale=0.3]{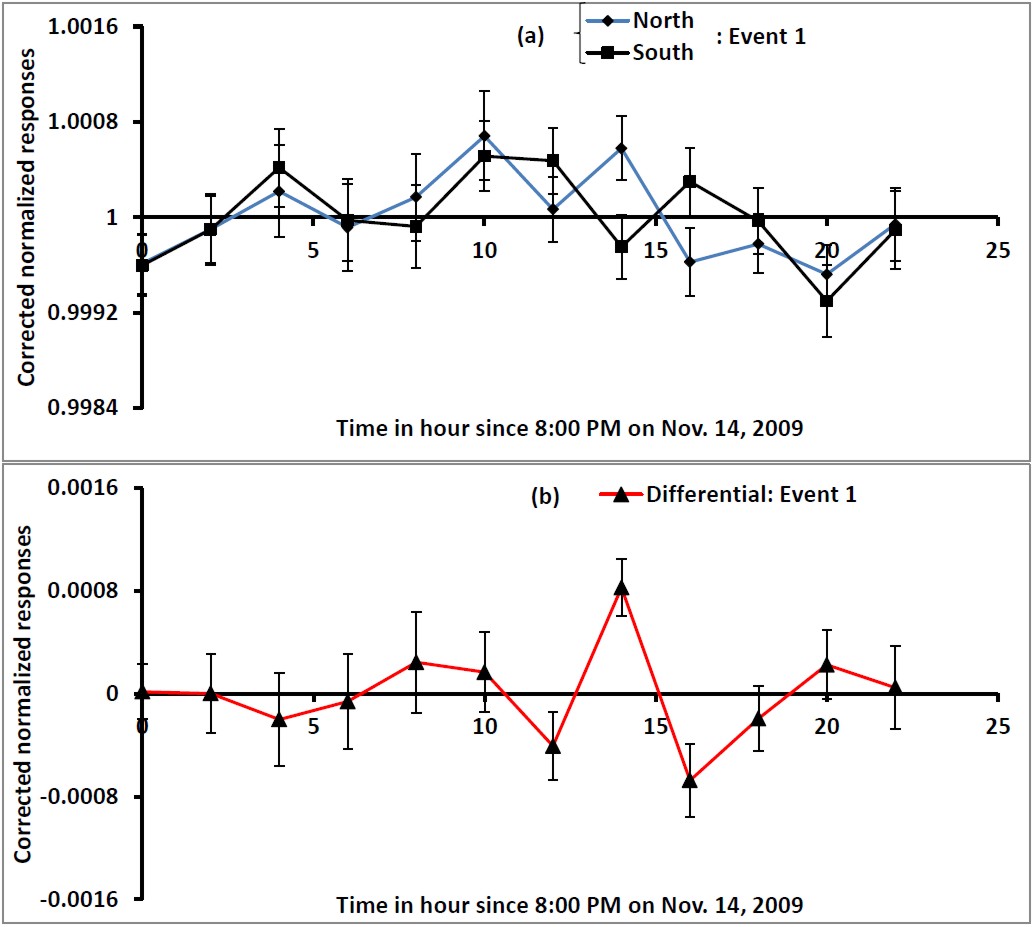}}
\captionof{figure}{Corrected results and their standard errors of (a) the individual (N, S) and (b) the differential responses, for Event 1.}
\label{figure:conindr1611}
\end{minipage}
\\ 
\noindent%
\begin{minipage}{\linewidth}
\makebox[\linewidth]{
  \includegraphics[keepaspectratio=true,scale=0.3]{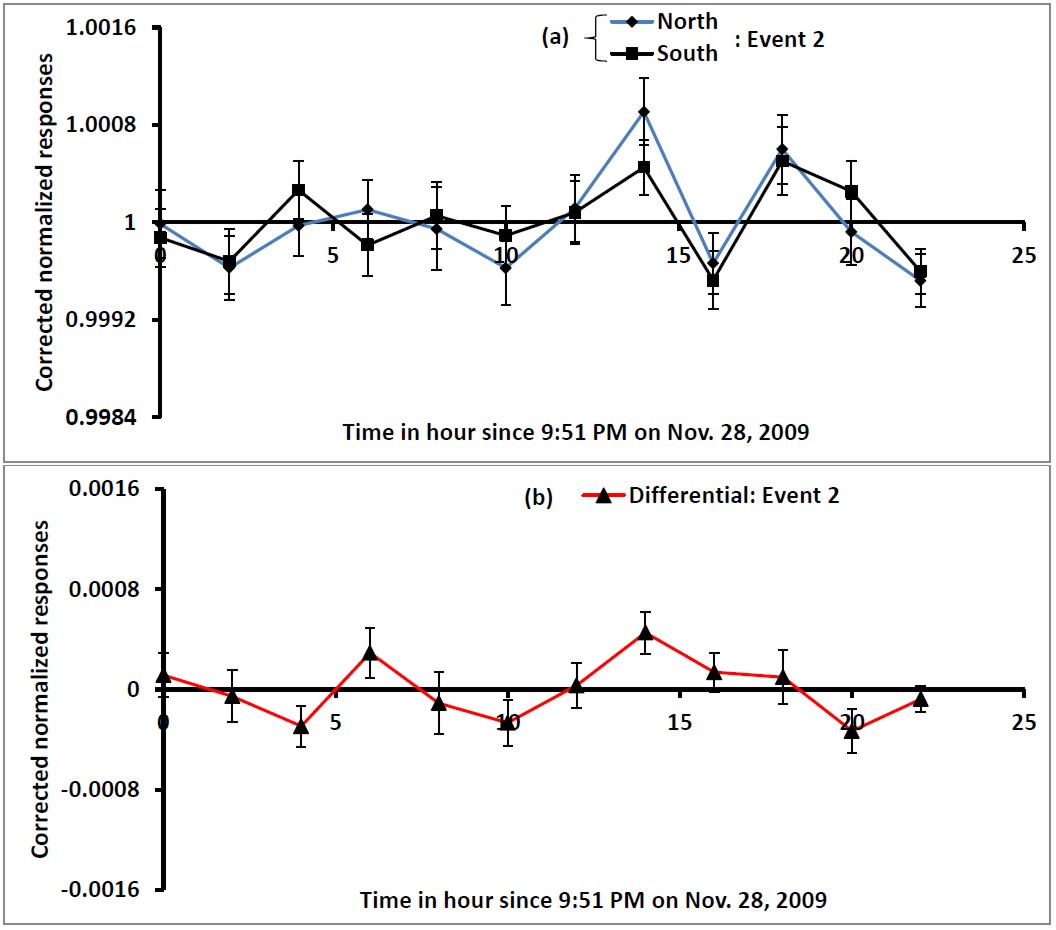}}
\captionof{figure}{Corrected results and their standard errors of (a) the individual (N, S) and (b) the differential responses, for Event 2.}
\label{figure:conindr2612}
\end{minipage}
\\ \\
\subsection{Analysis using least-squares fitting}
\label{sec:lsfncr} 
Following French \cite{FR147} and Jenkins and White \cite{JE1}, and also emphasized by Marinov \cite{MA21, MA22, MA23}, any diurnal regularity of variation can be approximated as a sinusoidal variation. We can fit by weighted least-squares the following equation to the differential responses.
\begin{equation}
\label{eq:SRviolatonmodel} 
R_{diff\_fit}=A \sin (\omega t+\delta )
\end{equation}
where $R_{diff\_fit}$ is a differential responses, $A$ is the amplitude, $ \delta $ is the phase and $ \omega =\frac{2\pi}{23~hr~56~min}  $ is the sidereal frequency. Note that the sidereal frequency has been used to represent azimuth change to represent all direction.   \\ \\
Using the differential-responses for this fitting, the fit for Event 1 is
\begin{equation}
\label{eq:modelev1} 
\hat{Y}_{1}=4.494\times 10^{-5}   \sin (\omega t+1.699621\pi)                       
\end{equation}
for Event 2 is
\begin{equation}
\label{eq:modelev2} 
\hat{Y}_{2}=1.037\times 10^{-4}   \sin (\omega t+1.22847\pi)                       
\end{equation}
The plots of the fits to (\ref{eq:SRviolatonmodel}) as shown in Eqs. (\ref{eq:modelev1}) and (\ref{eq:modelev2}) as well as the original data are plotted in Figs. \ref{figure:fit1615} (a) and (b) for Events 1 and 2, respectively. \\ \\
\noindent%
\begin{minipage}{\linewidth}
\makebox[\linewidth]{
  \includegraphics[keepaspectratio=true,scale=0.3]{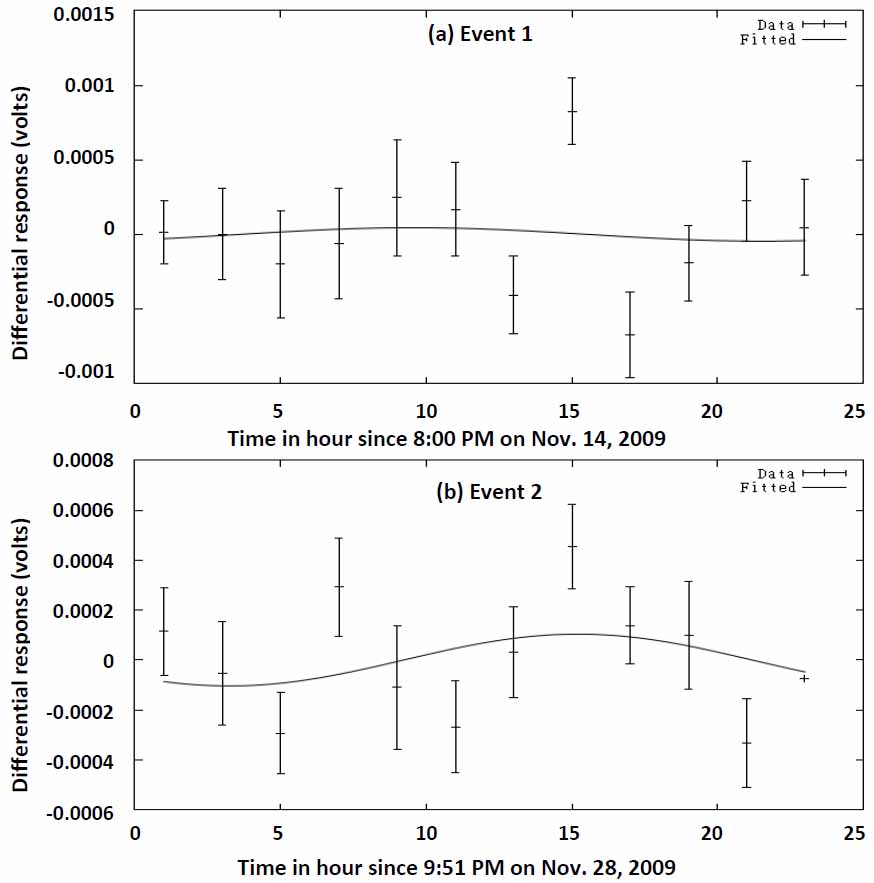}}
\captionof{figure}{The data (with standard errors) and the model (\ref{eq:SRviolatonmodel}) fitting. The fitted parameters are: (a) $A=(4.49 \pm 16.72) \times 10^{-5}$ and $ \delta=(1.7 \pm 1.2)\pi $ with the coefficient of determination of $R^{2}\cong 0.008$ for Event 1, (b) $ A=(1.04 \pm 1.02)\times 10^{-4} $ and $\delta  = (1.22 \pm 0.31)\pi  $ with the coefficient of determination of $R^{2}\cong 0.1$ for Event 2.}
\label{figure:fit1615}
\end{minipage}
\\ \\ \\
Using Eqs. (\ref{eq:northnormalized3}) and (\ref{eq:SRviolatonmodel}), we can write the following approximate relation:
\begin{equation}
\label{eq:limit} 
v_{limit}=\frac{1}{2}A\times c_{0}
\end{equation}
where $ v_{limit} $ is the limit of the variation of speed of light due to the motion of the Earth, $ c_{0} $ is the speed of light in vacuum and $A$ is the estimated amplitude.  \\ \\
Therefore, using the estimated amplitudes as shown in Figs. \ref{figure:fit1615} (a) and (b) (also, in Eqs. (\ref{eq:modelev1}) and (\ref{eq:modelev2})) and the value of the speed of light in vacuum $ (3\times10^{5}) $ km/sec in Eq. (\ref{eq:limit}), we calculate the maximum limit of the speed of the Earth in terms of the speed of light relative with a preferred frame (denoted by "limit" as follows).:   
\begin{enumerate}
\item The limit based on event 1: $ \leq (7\pm 25) $ km/sec 
\item The limit based on event 2: $ \leq (15\pm 15) $ km/sec 
\end{enumerate}
The fit, as shown in Fig. \ref{figure:fit1615}, with the coefficient of determination of $R^{2}< 0.1$ does not assure us that the theoretical model (\ref{eq:SRviolatonmodel}) for the present experiment based on a Galilean transformation with the sidereal frequency $ \omega =\frac{2\pi}{23~hr~56~min}  $ is correct. \\ \\
We can learn more about the fitting by using some quantitative approaches. These include certain hypothesis tests based on the $ t $-statistic and Analysis of Variance (ANOVA) on the regression coefficient $A$ in Eq. (\ref{eq:SRviolatonmodel}). The null hypothesis is that the parameter $ ~A=0 $. We accept the null hypothesis at $95\%$ confidence level and conclude that there is no sinusoidal relationship with the sidereal frequency ($ \omega=\frac{2\pi}{23~hr~56~min} $) based on the data of the corrected-differential responses.    \\ \\
The above discussion indicates that there is no physical meaning of the model in Eq. (\ref{eq:SRviolatonmodel}) according to a Galilean transformation (violation of SR) based on the data of the present measurements. However, we need to make a comparison of our outcome with other experimental investigations as well as to know the sensitivity of our apparatus by comparing it with the known value of the speed of Earth with respect to a preferred frame of reference. Therefore, we use the value $ (7\pm 25) $ km/s as the speed of the Earth relative to a preferred frame of reference to perform comparisons in Section \ref{sec:summaryoutcome}.     
\subsection{Fourier analysis}
We perform Fourier analysis to decompose the differential responses into its frequency components to look for any harmonic that presents identical frequencies according to our expectation in this work. Following Jenkins and Watts \cite{JE34}, we perform Fourier decomposition of the mean square for the differential data. These spectrum analysis are presented in Figs. \ref{figure:Specevent1} and \ref{figure:Specevent2}.   \\ \\
\noindent%
\begin{minipage}{\linewidth}
\makebox[\linewidth]{%
  \includegraphics[keepaspectratio=true,scale=0.35]{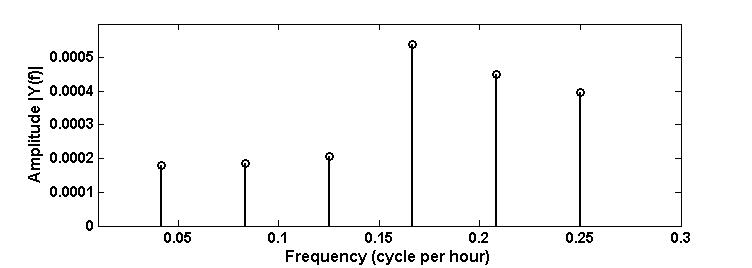}}
\captionof{figure}{A frequency spectrum of differential signals for Event 1.}
\label{figure:Specevent1}
\end{minipage}
\\ \\
\noindent%
\begin{minipage}{\linewidth}
\makebox[\linewidth]{%
  \includegraphics[keepaspectratio=true,scale=0.35]{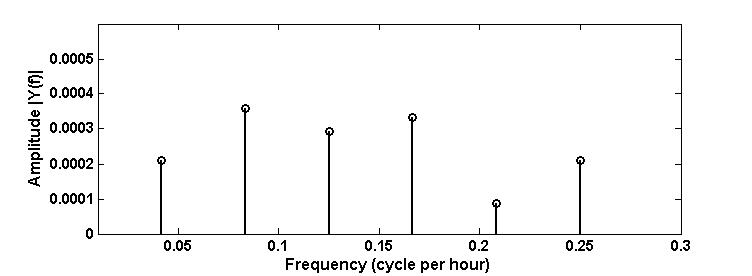}}
\captionof{figure}{A frequency spectrum of differential signals for Event 2.}
\label{figure:Specevent2}
\end{minipage}
\\ \\
The random values, as illustrated in Figs. \ref{figure:Specevent1} and \ref{figure:Specevent2}, do not present any physical meanings according to our assumption in this work where harmonics with the frequency of approximately $ \frac{1}{24} $ cycles/hour is our expectation if the speed of light is anisotropic. However, we noted about the errors as well as the infra-red radiation could be responsible for these disturbances. This outcome can be interpreted using Einstein's special theory of relativity. 
\subsection{Summary of the present outcomes}
\label{sec:summaryoutcome} 
All final outcomes are summarized in Table \ref{table:summary72}. In order to understand the sensitivity of the present test, we follow the method in the reports \cite{PA2, SH36} by using the $ \text{Ratio}=\frac{\text{Expected ~limit}}{\text{Observed ~limit}} $ where we consider different preferred frames of reference from classical (relative with the center of the Sun) to the modern (relative with the CMB). Our experimental observations, as presented in Table \ref{table:summary72}, are consistent with the special theory of relativity.   \\ 
\begin{table}[ht]
\small
      \centering
\caption{Summary of the outcomes of the present experiment.}
\label{table:summary72}
\begin{tabular}{ |l|l|l||l| }
\hline
\textbf{Tools} & \textbf{Observed} & \textbf{Expected} & \textbf{Comments} \\
\hline
\multicolumn{4}{ |c| }{\emph{\textbf{If SR invalid}} } \\
\hline
Spectral analysis for   & No significant & Maximum & Inconsistent \\
 the 1 $cpd$ frequency& contribution & contribution & \\
\hline
\multirow{2}{*}{Variation in the speed of light} & $ (7\pm 25) $ km/s in Event 1 & \multirow{2}{*}{$ \geq $371 km/s \cite{SM113}} & \multirow{2}{*}{Inconsistent}\\
 &$ (15\pm 15) $ km/s in Event 2 &  & \\ \hline
 Shape of & Same variation in Event 1 & Opposite & Inconsistent \\
 individual responses & Same variation in Event 2 & variation (Fig. \ref{fig:sriv310} ) & 
 Inconsistent \\ \hline \hline
\multicolumn{4}{ |c| }{\emph{\textbf{Sensitivity=Expected/Observed}}} \\
\hline
The CMB &$ (7\pm 25) $ km/s & 371 km/s \cite{SM113} & 53 \\ \hline
The Sun & $ (7\pm 25) $ km/s & 30 km/s & 4 \\ \hline
Marinov \cite{MA21, MA22, MA23} &$ (7\pm 25) $ km/s & $ (360\pm 40) $ km/s & 51 \\ 
\hline
\hline
\end{tabular}
\end{table}
\section{Conclusion}
It is well known that D.C. Miller \cite{MI35, SH36} devoted himself trying to find the absolute motion of the Earth by experimental observations. With significant improvements of his apparatus, D.C. Miller claimed \cite{MI35} to have found the long-sought absolute motion of the Earth in terms of the speed of light. However, in 1955, the year that Albert Einstein died, Shankland \textit{et al.} \cite{SH36} in Cleveland published an elaborate statistical analysis of Miller's work, judging his anomalous, small but positive results to have been caused by errors. This discussion indicates the necessity of a rigorous statistical analysis regardless of what the improvements are in the apparatus.  \\ \\
We performed a rigorous statistical analysis where our prime objective is to identify the trends due to the disturbances, and then eliminate them from the data, we are left with fluctuating series which may be, at one extreme, purely random (if SR is valid), or at the other, a smooth oscillating movement with peaks and troughs occur at equal intervals of time (for present work 12 hours) if SR is invalid. \\ \\
Our results do not report any significant diurnal variations as one may expect according to the movements of the Earth. Consequently, the limit is insignificant. This is consistent with Einstein's Special Relativity.    
\section*{Acknowledgements}
We are indebted to Dr. Matthew George for numerous discussions, and thank Dr. S. Sargoytchev, Dr. Eamonn McKernan and Nick Balaskas for help with the setup. This work is supported in part by the National Science and Engineering Research Council, Thoth Technology Inc. and York University, Canada.      

\end{document}